\documentclass[twocolumn, switch]{article} 

\usepackage{preprint}

\usepackage{amsmath,amsfonts}
\usepackage{algorithmic}
\usepackage{algorithm}
\usepackage{array}
\usepackage{textcomp}
\usepackage{stfloats}
\usepackage{url}
\usepackage{verbatim}

\usepackage{cite}
\usepackage{subcaption}
\usepackage{multirow}
\usepackage{color}
\usepackage{booktabs}
\usepackage[pagebackref,breaklinks,colorlinks,bookmarks=false]{hyperref}
\usepackage{graphicx}
\usepackage{titlesec}
\titlespacing\section{0pt}{12pt plus 3pt minus 3pt}{1pt plus 1pt minus 1pt}
\titlespacing\subsection{0pt}{10pt plus 3pt minus 3pt}{1pt plus 1pt minus 1pt}
\titlespacing\subsubsection{0pt}{8pt plus 3pt minus 3pt}{1pt plus 1pt minus 1pt}

\title{Dynamic Kernel-Based Adaptive Spatial Aggregation for Learned Image Compression}

\usepackage{authblk}

\author[1]{Huairui Wang}
\author[1]{Nianxiang Fu}
\author[1*]{Zhenzhong Chen}
\author[2]{Shan Liu}
\affil[1]{School of Remote Sensing and Information Engineering, Wuhan University}
\affil[2]{Tecent America}

\begin{document}

\twocolumn[ 
  \begin{@twocolumnfalse} 
  
\maketitle

\begin{abstract}

Learned image compression methods have shown superior rate-distortion performance and remarkable potential compared to traditional compression methods. Most existing learned approaches use stacked convolution or window-based self-attention for transform coding, which aggregate spatial information in a fixed range. In this paper, we focus on extending spatial aggregation capability and propose a dynamic kernel-based transform coding. The proposed adaptive aggregation generates kernel offsets to capture valid information in the content-conditioned range to help transform. With the adaptive aggregation strategy and the sharing weights mechanism, our method can achieve promising transform capability with acceptable model complexity. Besides, according to the recent progress of entropy model, we define a generalized coarse-to-fine entropy model, considering the coarse global context, the channel-wise, and the spatial context. Based on it, we introduce dynamic kernel in hyper-prior to generate more expressive global context. Furthermore, we propose an asymmetric spatial-channel entropy model according to the investigation of the spatial characteristics of the grouped latents. The asymmetric entropy model aims to reduce statistical redundancy while maintaining coding efficiency. Experimental results demonstrate that our method achieves superior rate-distortion performance on three benchmarks compared to the state-of-the-art learning-based methods.

\end{abstract}

\vspace{0.4cm}

  \end{@twocolumnfalse} 
] 

\newcommand\blfootnote[1]{%
\begingroup
\renewcommand\thefootnote{}\footnote{#1}%
\addtocounter{footnote}{-1}%
\endgroup
}

\section{INTRODUCTION}

{\blfootnote{Corresponding author: Zhenzhong Chen, Email:zzchen@ieee.org}}

Image compression has been a fundamental problem in image and video processing for decades. It plays a critical role in image and video transmission and storage, especially for applications with limited bandwidth and storage capacities, such as mobile devices and cloud storage. In the past, most image compression methods were designed using handcrafted algorithms, such as JPEG \cite{wallace1992jpeg}, BPG \cite{bellard2014bpg}, and VVC \cite{vvc}. These methods typically use a block-based transform coding framework and rely on human-crafted features and heuristics to remove the redundancy in the image. Although these methods have achieved good compression performance, their coding efficiency is limited by the separately optimized framework and the need for more flexibility to adapt to different image contents.
\begin{figure}[t]
	\centering
	\includegraphics[width=0.47\textwidth]{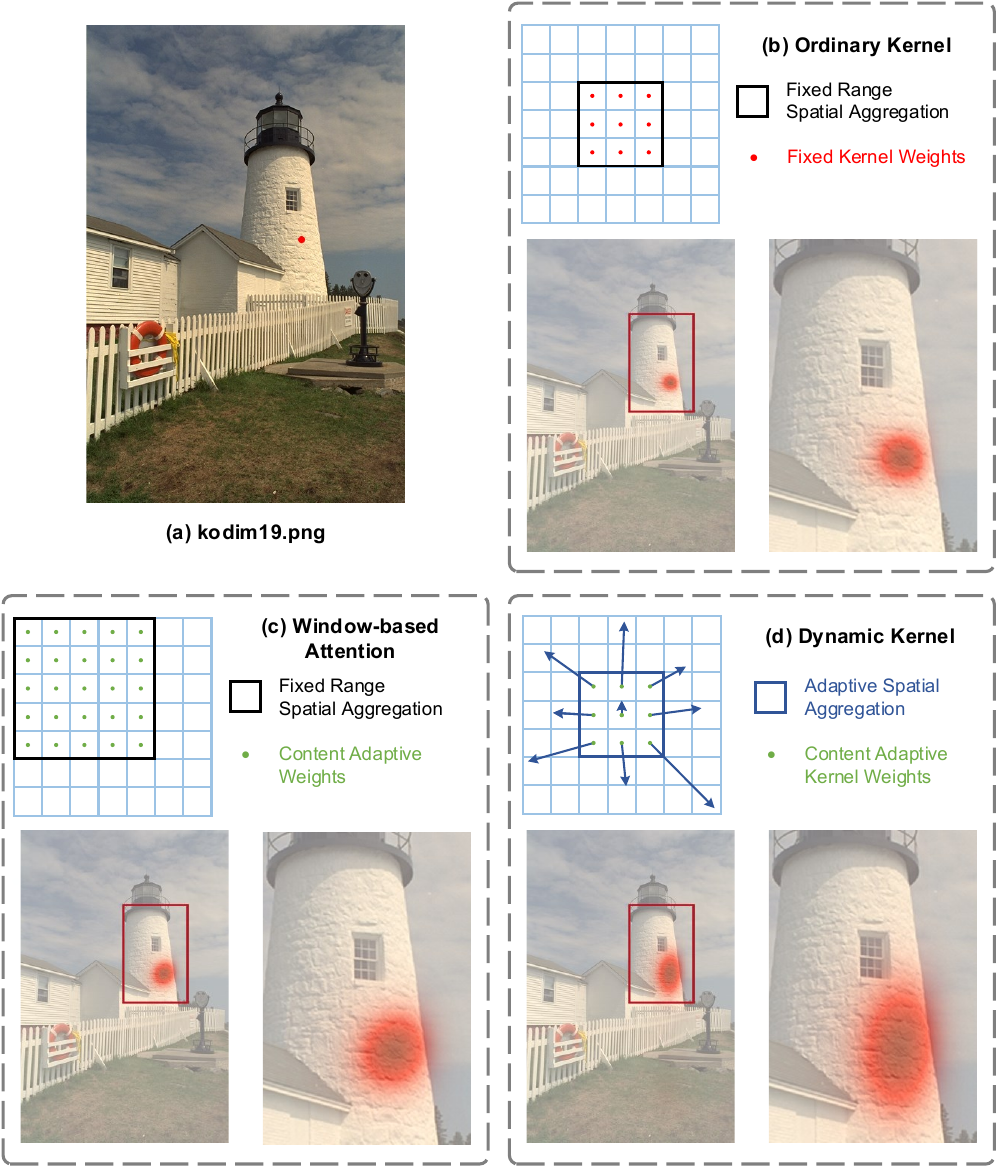}
	\caption{Characteristic differences between the ordinary kernel, window-based attention, and our dynamic kernel. For the visualization of the effective receptive fields, we choose Cheng2020\cite{cheng2020learned} and STF\cite{zou2022devil} as the representative methods using ordinary kernel and window-based attention, respectively. \textcolor{red}{The red point} in (a) denotes the target point.}
	\label{fig:dynamic_kernel}
\end{figure}

Recently, deep learning has demonstrated remarkable performance in many computer vision tasks, such as image recognition and object detection \cite{he2016deep, Dai:2017, liu2021swin}. Inspired by the success of deep learning, there has been increasing interest in applying deep learning techniques to image compression \cite{balle2016end, balle2018variational, minnen2018joint, cheng2020learned, zhu2022transformerbased, he2022elic}. In particular, recent works have explored the use of deep neural networks for image compression by training a neural network to learn the image compression process end-to-end. The end-to-end optimized manner is attractive because it can automatically learn the most suitable compression parameters for different types of images and metrics. It has the potential to outperform traditional compression methods by exploiting the full capacity of deep neural networks. Some recent learned image compression (LIC) approaches \cite{zou2022devil,zhu2022transformerbased,xie2021enhanced, chen2022two,he2022elic,wang2022neural} have outperformed VTM \cite{vvc} which is the latest traditional coding standard reference software, in terms of PSNR and MS-SSIM. Furthermore, learning-based compression methods have more promising potential for improving perceptual and subjective quality \cite{mentzer2020high} since they can adjust optimizing direction by replacing loss functions. These suggest that LIC has a great capacity for developing the next-generation image compression framework. 

Recently, numerous LIC methods \cite{liu2019non,cheng2020learned,wu2021learned, guo2021causal, he2022elic} enhance the variational auto-encoder architecture \cite{balle2018variational} with CNN-based transform coding. For example. Cheng \emph{et al.} \cite{cheng2020learned} introduced residual blocks to implement transformation between images and latents. CNN-based methods tend to stack convolutional layers and aggregate information in a fixed range according to kernel size. Besides, due to the inductive bias of regular convolution, it has translation equivariance, and its kernel weights are fixed during inference. In addition, since Vision Transformers \cite{dosovitskiy2020image, liu2021swin} have achieved eye-catching performance in many visual tasks, several image compression methods \cite{lu2021transformer, zou2022devil, zhu2022transformerbased} utilized window-based attention from Swin Transformer \cite{liu2021swin} to perform spatial information aggregation. The window-based attention can generate adaptive weights based on inter-token correlation within a preset size window. In this paper, we focus on enhancing the spatial aggregation capacity in a dynamic range and further boosting Rate-Distortion (RD) performance. Deformable convolution (DCN) \cite{Dai:2017, zhu2019deformable} fits well with adaptive spatial aggregation, but its high computation complexity and GPU memory cost hinder its use in transform coding. To this end, we introduce Lite DCN (LDCN) from large-scale model \cite{wang2022internimage} in our framework. LDCN, named dynamic kernel in this paper, can generate learned offsets and group-sharing weights to break the limitation of fixed range spatial aggregation. With the content-adaptive kernel offsets and weights, it is easy for the model to aggregate spatial information in a dynamic range. Figure~\ref{fig:dynamic_kernel} illustrates the main differences between the mainstream operations and the dynamic kernel. We provide the visualization of the effective receptive fields \cite{gu2021interpreting} of Cheng2020\cite{cheng2020learned}, STF\cite{zou2022devil}, and our method. The visualization can demonstrate that dynamic kernel-based transform coding can aggregate information from a more extensive and content-conditioned range.

As another core component of LIC, the entropy models are designed for distribution parameter estimation and probabilistic prediction. Ball\'{e} \emph{et al.} \cite{balle2018variational} proposed a hyper-prior as side information to store global context. Minnen \emph{et al.} \cite{minnen2018joint} utilized the spatial autoregressive model to capture context from decoded latents. However, this kind of context model cannot perform parallel computation, so later they \cite{minnen2020channel} proposed a channel-wise autoregressive entropy model by splitting the latent into slices and sequentially encoding them. The decoded slices can participate in encoding the remaining slices as channel-wise context. He \emph{et al.} \cite{he2022elic} investigated energy distribution and extended the model to unevenly grouped channel-wise context. In this paper, we first define a generalized efficient entropy model using a coarse-to-fine context. Then we introduce the above-mentioned dynamic kernel in hyper-prior to generate more expressive global context. Moreover, we analyze the energy response and the spatial correlation in the latents and propose the Asymmetric Spatial-channel Entropy Model to promote RD performance while keeping a satisfying inference speed. The main contributions of this paper can be summarized as follows:
\begin{itemize}
	\item We introduce the dynamic kernel-based adaptive spatial aggregation for transform coding. With the spatial aggregation assisted by offsets and the sharing weights, the transform module effectively reduces the bit-rate while maintaining computation complexity.
	\item A generalized coarse-to-fine entropy model is developed considering the global, channel-wise, and spatial context. Based on the model, we introduce a dynamic kernel into hyper-prior to generate a powerful global context. Besides, we further investigate the latent distribution and spatial correlation and propose an asymmetric spatial-channel entropy model to remove statistical redundancy efficiently.
	\item With the merit of the adaptive spatial aggregation and the asymmetric entropy model, our method surpass VTM-12.1 and other state-of-the-art methods on three benchmarks. Extensive experiments demonstrate that the proposed modules can effectively improve RD performance while maintaining model complexity.
\end{itemize}

The remainder of this paper is organized as follows. Section \uppercase\expandafter{\romannumeral2} introduces the related work. In Section \uppercase\expandafter{\romannumeral3}, we present the overall framework DKIC with  detailed  discussions about Dynamic Kernel and Asymmetric Spatial-channel Entropy Model. In Section \uppercase\expandafter{\romannumeral4}, we show the performance of DKIC and discuss the importance of each module, and then we conclude in Section \uppercase\expandafter{\romannumeral5}.

\section{Related Work}
\subsection{Learned image compression} 
The pioneering work of Ball\'{e} \emph{et al.} \cite{balle2016end} firstly proposed a CNN-based learned image compression model which uses stacked convolutions and generalized divisive normalization (GDN) layers to achieve transform coding. Then Ball\'{e} \emph{et al.} \cite{balle2018variational} modeled the image compression framework as a variational auto-encoder and proposed a hyper-prior as side information. Following the progress in probabilistic generative models, Minnen \emph{et al.} \cite{minnen2018joint} proposed spatial autoregressive priors and used masked convolution for sequential coding and decoding with decoded latents. Mishra \emph{et al.} \cite{9144534} proposed a a Wavelet-based Deep Auto Encoder-Decoder Network based image compression. Cheng \emph{et al.} \cite{cheng2020learned} presented a residual block-based transform coding, and they also introduced Discretized Gaussian Mixture Model (GMM) to model the distribution of the latent representation. Cai \emph{et al.} \cite{8531758} proposed a CNN-based multi-scale decomposition transform and content adaptive rate allocation to achieve rariable rate compression. He \emph{et al.} \cite{he2022elic} experimentally proved that stacking the residual blocks as a nonlinear transform can achieve promising rate-distortion performance, even without GDN layers. Tang \emph{et al.} \cite{9858899} proposed a compression method by integrating graph attention and asymmetric convolutional neural network, achieving promising rate-distortion performance.

Besides, inspired by the success of Transformer architectures in natural language processing, Transformers have been introduced to the vision domain and have shown performance competitive with CNN in many vision tasks, including image classification \cite{liu2021swin, liu2022swin}, object detection \cite{liu2022swin} and image restoration \cite{liang2021swinir}. Motivated by those works, Transformer-based models have been explored for learned image compression. For instance, Lu \emph{et al.} \cite{lu2021transformer} proposed a Transformer-based image compression method that stacks Swin-Transformer block and convolutions to improve the information embedding ability of the network. Zhu \emph{et al.} \cite{zhu2022transformerbased} also proposed Swin-Transformer-based nonlinear transforms for image and video compression. Zou \emph{et al.} \cite{zou2022devil} proposed a Symmetrical Transformer Framework with window-based attention to capturing correlations among spatially neighboring elements. Kim \emph{et al.} \cite{kim2022joint} proposed an Information Transformer to generate global and local compression priors.

\subsection{Entropy models}
The entropy model is a crucial component for estimating the distribution of discrete latent representations. In the field of learned image compression, the entropy model was first proposed by Ball\'{e} \emph{et al.} \cite{balle2018variational}, in which a hyper-prior was incorporated to capture global spatial dependencies in latent representations. In this model, each latent element is modeled as a zero-mean Gaussian with a standard deviation conditioned on the hyper-prior. A non-parametric fully factorized density model is used to model the side information. In subsequent work, such as \cite{minnen2018joint, cheng2020learned}, hyper-priors and spatial decoded latents were jointly utilized for context modeling in an autoregressive manner, resulting in improved image compression performance. However, this approach can be time-consuming due to the serial processing steps. To this end, He \emph{et al.} \cite{he2021checkerboard} investigated the spatial characteristics of the latents and proposed parallelization-friendly checkerboard context. In the meantime, Minnen \emph{et al.} \cite{minnen2020channel} designed a channel-wise entropy model to reduce statistical redundancy along the channel dimension. He \emph{et al.} \cite{he2022elic} proposed an uneven grouping strategy to develop a spatial-channel contextual adaptive model, further enhancing the coding performance. Recently, Lu \emph{et al.} \cite{lu2022high} and Lin \emph{et al.} \cite{lin2023multistage} explored the feasibility of a multi-stage contextual model.

\subsection{Dynamic kernel}
Deformable convolution (DCN) is a representative method in dynamic kernels. It is a type of convolutional operation that enables neural networks to have more flexible receptive fields, which is widely used in various tasks, including object detection \cite{Dai:2017, zhu2019deformable}, image classification \cite{zhu2019deformable} and video super-resolution \cite{wang2019edvr, Chan_2022_CVPR}. The concept of DCN was first proposed by Dai \emph{et al.} \cite{Dai:2017}, in which the network can learn additional offsets to gather information from beyond its regular local neighborhood. Zhu \emph{et al.} \cite{zhu2019deformable} introduced DCNv2, which includes a modulation mechanism that not only learns the offset for each sample but also modulates the learned feature amplitude. This modulation mechanism enables the network module to vary both the spatial distribution and the relative influence of its samples. Furthermore, Wang \emph{et al.} \cite{wang2022internimage} extended DCNv2 to DCNv3, which introduces a multi-group mechanism, shares weights among convolutional neurons, and normalizes modulation scalars along sampling points. Moreover, Yang \emph{et al.} \cite{yang2019condconv} introduced conditionally parameterized convolutions which learn specialized convolutional kernels for the input feature. Chen \emph{et al.} \cite{chen2020dynamic} proposed a dynamic convolution that can dynamically aggregate multiple parallel convolutional kernels based on their attention weights.

\section{The Proposed Method}
\subsection{Framework Description}\label{framework}
Ball\'{e} \emph{et al.} \cite{balle2016end, balle2018variational} modeled the image compression framework as a variational auto-encoder and optimized the networks in an end-to-end fashion. Since then, most LIC frameworks have followed the paradigm and improved the compression performance rapidly. Based on the recent representative work \cite{minnen2018joint, minnen2020channel, he2022elic} and the transform coding theory \cite{goyal2001theoretical}, we formulate a generalized image compression framework as follows:
\begin{equation}
	\begin{aligned}
		\boldsymbol{y}&=g_a(\boldsymbol{x};\boldsymbol{\phi}),\\
		\boldsymbol{\hat y}&=Q(\boldsymbol{y}-\mu)+\mu,\\
		\boldsymbol{\hat x}&=g_s(\boldsymbol{\hat y};\boldsymbol{\theta})\\
	\end{aligned}
	\label{eq:video_compression}
\end{equation}
where $\boldsymbol{x}$ and $\boldsymbol{\hat{x}}$ denote the input and the decompressed images. $g_a$ and $g_s$ are analysis and synthesis transforms. $\boldsymbol{\phi}$ and $\boldsymbol{\theta}$ are learned parameters of the analysis and the synthesis transform networks, respectively. $Q$ represents quantization operation. 

During encoding, we send the input $\boldsymbol{x}$ to the encoder $g_a$ with learned parameters $\boldsymbol{\phi}$, then we get the latent representation $\boldsymbol{y}$. We follow the previous work \cite{minnen2018joint, minnen2020channel} and encode the quantized residual between $\boldsymbol{y}$ and $\mu$ to the bitstream. To encode $\boldsymbol{y}-\mu$ with arithmetic coding, we quantize it by $Q$. Then, in the decoding process, the decoder $g_s$ will transform the decoded latents $\boldsymbol{\hat y}$ into the reconstructed images. 
\begin{figure*}[t]
	\centering
	\includegraphics[width=0.92\textwidth]{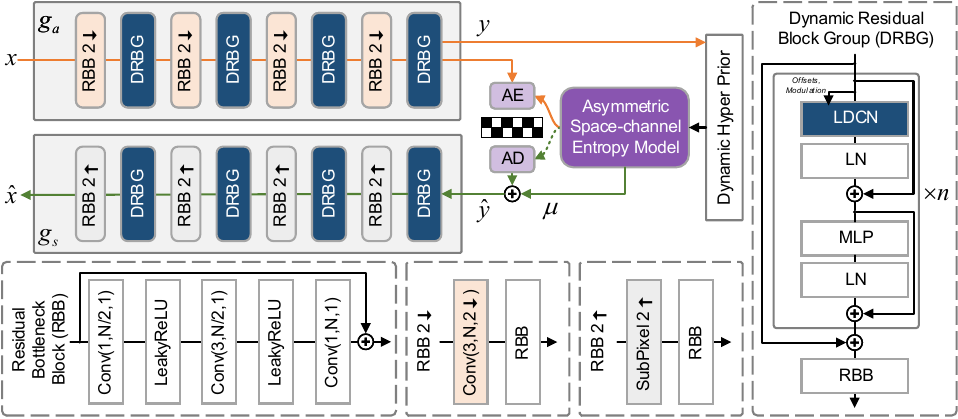}
	\caption{Overview of our proposed image compression framework DKIC. We use $g_a$ to transform the input $x$ into latent representation $y$, and propose Asymmetric Space-channel Entropy Model to estimate the distribution parameters ($\mu$ and $\sigma$) of $y$. Following \cite{minnen2018joint}, we quantize and compress $y-\mu$ to the bitstream. After entropy decoding, we restore the image $\hat{x}$ from $\hat{y}$ with inverse transform network $g_s$.}
	\label{fig:overview}
\end{figure*}

Following the previous work \cite{balle2018variational, minnen2018joint}, we use a single Gaussian distribution to model each symbol, so the variances $\mu$ and the means $\sigma$ of symbols are estimated by the entropy model. For building a comprehensive entropy model in a coarse-to-fine manner, we introduce the hyper-prior \cite{balle2018variational}, the channel-wise autoregressive context \cite{minnen2020channel}, and the spatial context entropy model \cite{he2021checkerboard, he2022elic} in the framework. Specifically, the generalized entropy model encodes the $\boldsymbol{y}$ into side information $\boldsymbol{z}$ using hyper-prior network \cite{balle2018variational}. Then the entropy model splits $\boldsymbol{y}$ into $N$ slices (${\boldsymbol{y}^1,\boldsymbol{y}^2,..., \boldsymbol{y}^N }$) along the channel dimension, and encodes the slices in an autoregressive way. Within the parameter estimation of each slice, the previously decoded slices can assist in estimating the distribution of the current slice. Besides, each slice ($i$-th slice for example) is also divided into $K$ parts in the spatial dimension (${\boldsymbol{y}^i_1, \boldsymbol{y}^i_2,..., \boldsymbol{y}^i_K }$). The entire entropy model can be formulated as:
\begin{equation}
	\begin{aligned}
		\boldsymbol{z}=&h_a(\boldsymbol{y}; \boldsymbol{\phi_h}),
		\boldsymbol{\hat z}=Q(\boldsymbol{z})\\
		\boldsymbol{gc} =& h_s(\boldsymbol{\hat z};\boldsymbol{\theta_h})\\
		\boldsymbol{cc} =& g_{cp}(\boldsymbol{\hat z},\boldsymbol{\hat y}^{< i};\boldsymbol{\theta_{cp}}), 1 \le i \le N\\
		\boldsymbol{sc} =& g_{sp}(\boldsymbol{\hat z},\boldsymbol{\hat y}^i_{< j};\boldsymbol{\theta_{sp}}), 1 \le j \le K\\
		\mu^i_j, \sigma^i_j =&g_{ep}(\boldsymbol{gc},\boldsymbol{cc},\boldsymbol{sc};\boldsymbol{\theta_{ep}}),\\
		p_{\boldsymbol{\hat y}}(\boldsymbol{\hat y}^i_j| \boldsymbol{ctx})=&[\mathcal N(\mu^i_j,(\sigma^i_j)^2)]*\mathcal U(-\frac{1}{2},\frac{1}{2})(\boldsymbol{\hat y}^i_j)\\
		with \quad  \boldsymbol{ctx} =& (\boldsymbol{\hat z},\boldsymbol{\hat y}^{< i},\boldsymbol{\hat y}^i_{< j},\boldsymbol{\theta_h},\boldsymbol{\theta_{cc}},\boldsymbol{\theta_{sc}},\boldsymbol{\theta_{ep}})
	\end{aligned}
	\label{eq:general_EM}
\end{equation}
where $h_a$, $h_s$, $\boldsymbol{\phi_h}$ and $\boldsymbol{\theta_h}$ are the hyper analysis and synthesis transform and their learnable parameters in the hyper-prior. $\boldsymbol{z}$ and $\boldsymbol{\hat z}$ stand for the side information and its quantized counterpart. $g_{cp}$, $g_{sp}$, $g_{ep}$, $\boldsymbol{\phi_{cp}}$, $\boldsymbol{\phi_{sp}}$ and $\boldsymbol{\theta_{ep}}$ are the channel, spatial and final parameter prediction networks and their learnable parameters. $\boldsymbol{\hat z}$ is fed to the hyper synthesis transform $h_s$ to obtain the global coarse context $\boldsymbol{gc}$. Then we use $\boldsymbol{gc}$ and the already decoded slices (${\boldsymbol{y}^1,\boldsymbol{y}^2,..., \boldsymbol{y}^{i-1} }$) to generate channel-wise context $\boldsymbol{cc}$ with the network $g_{cp}(\cdot)$. In the mean time, the entropy model generates the spatial context $\boldsymbol{sc}$ using the decoded parts (${\boldsymbol{y}^i_1, \boldsymbol{y}^i_2,..., \boldsymbol{y}^i_{j-1} }$). Finally, the Gaussian parameters $\mu^i_j$ and $\sigma^i_j$ of the target position are predicted by the final parameter prediction network $g_{ep}$ (the \emph{Entropy Parameters} network in \cite{minnen2018joint}).

Since we provide a generalized entropy model, most existing entropy models can be seen as special cases. For example, when we set $K$ as the number of spatial feature points of $\boldsymbol{y}$ and remove the channel context, we get the spatial autoregressive prior in \cite{minnen2018joint}. When we set $N$ = 10 and divide $\boldsymbol{y}$ into 10 slices evenly along the channel dimension and remove the spatial context, we get the channel-wise autoregressive entropy model in \cite{minnen2020channel}. If we set $K$ as 2 and remove the channel context, we get the checkerboard context in \cite{he2021checkerboard}.

We enhance the framework defined above with dynamic spatial aggregation and asymmetric space-channel entropy model to improve the rate-distortion performance further. Specifically, we propose the dynamic kernel-based spatial aggregation to combine the advantages of the CNN-based \cite{minnen2018joint, cheng2020learned, he2021checkerboard, he2022elic} and Transformer-based \cite{lu2021transformer, zou2022devil, zhu2022transformerbased} methods and break through the limitations they share. A more detailed discussion will be presented in Section \ref{Dynamic_Kernel}. Moreover, we are inspired by the representative efficient entropy model \cite{balle2018variational, minnen2020channel, he2021checkerboard} and summarize the generalized coarse-to-fine entropy model as described in Equation \ref{eq:general_EM}. To enhance the expression of global context, we introduce the dynamic kernel into hyper-prior to aggregate global prior. Furthermore, based on the conclusion in \cite{he2022elic} and our investigation on spatial characteristics of latent representation, we propose the asymmetric spatial-channel entropy model to achieve high compression performance while maintaining efficiency. We introduce the proposed entropy model in Section \ref{ASCEM}.

\subsection{Dynamic Kernel for Image Compression}\label{Dynamic_Kernel}
Most recently learned image compression methods either use CNN \cite{cheng2020learned, he2021checkerboard, he2022elic} or Vision Transformer \cite{lu2021transformer, zou2022devil, zhu2022transformerbased} to transform the images into latent representation. CNN-based and Transformer-based compression methods can Gaussianize data effectively and achieve comparable rate-distortion performance to the traditional codec VTM, but they still have limitations. As shown in Figure\ref{fig:dynamic_kernel}, CNN-based methods stack ordinary convolution and aggregate spatial information in a fixed range according to their kernel size. Besides, due to the inductive bias of ordinary convolution, it has translation equivariance, and its kernel weights are fixed while processing different contents. Most Transformer-based compression methods use the window-based attention in Swin Transformer block \cite{liu2021swin} to achieve spatial information aggregation. With the merit of self-attention within a fixed-size window, this method can dynamically assign weights based on token similarity. However, even with the shifted window mechanism, they can only aggregate features within a fixed spatial range. 

To break through the above-mentioned limitations and further boost RD performance, we intend to introduce deformable kernel \cite{Dai:2017, zhu2019deformable} to achieve adaptive spatial aggregation. With the content-adaptive kernel offsets and weights in deformable convolution (DCN), it is easy for the kernel to aggregate spatial information in a dynamic range for efficient Gaussianization. In \cite{zhu2019deformable}, the DCN can be formulated as:
\begin{equation}
	\hat{F}(\mathbf{p})=\sum_{k=1}^Kw_{k}\cdot {F}(\mathbf{p}+\mathbf{p}_{k}+\Delta\mathbf{p}_{k}),
	\label{DCNv2}
\end{equation} 
where $w_{k}, \mathbf{p}_{k}, \Delta\mathbf{p}_{k}$ are the modulated kernel weights, the general sampling location and the additional learned offsets. $w_{k}$ is adaptive for each element of the input feature $F$. However, there is a consensus that deformable convolution consumes much more computation resources and GPU memory than ordinary convolution, so the blind introduction of deformable convolution may lead to the impracticality of the compression method. To preserve the advantages of deformable convolution while reducing its operational complexity, we introduce Lite DCN (LDCN) from InternImage\cite{wang2022internimage} as the dynamic kernel to aggregate information in a dynamic spatial range. Compared to the DCN in \cite{zhu2019deformable}, LDCN splits the input feature into groups and has the sharing modulated kernel weights within each group. Moreover, LDCN uses softmax normalization to replace the element-wise sigmoid normalization, alleviating the unstable gradient in training. The LDCN used in our method can be formulated as:
\begin{equation}
	\hat{F}(\mathbf{p})=\sum_{g=1}^G\sum_{k=1}^Kw_{gk}\cdot {F}(\mathbf{p}+\mathbf{p}_{k}+\Delta\mathbf{p}_{gk}),
	\label{LDCN}
\end{equation} 
where $G$ denotes the number of the split groups, $w_{gk}, \Delta\mathbf{p}_{gk}$ are the sharing modulated kernel weights and the additional learned offsets within $g$-th subgroup. Benefit from the group-wise sharing weights, the Lite DCN used in our method can be plugged into our method as dynamic transform operation without introducing heavy computation burden.
\begin{figure}[t]
	\centering
	\begin{subfigure}{.24\textwidth}
		\centering
		\includegraphics[width=\textwidth]{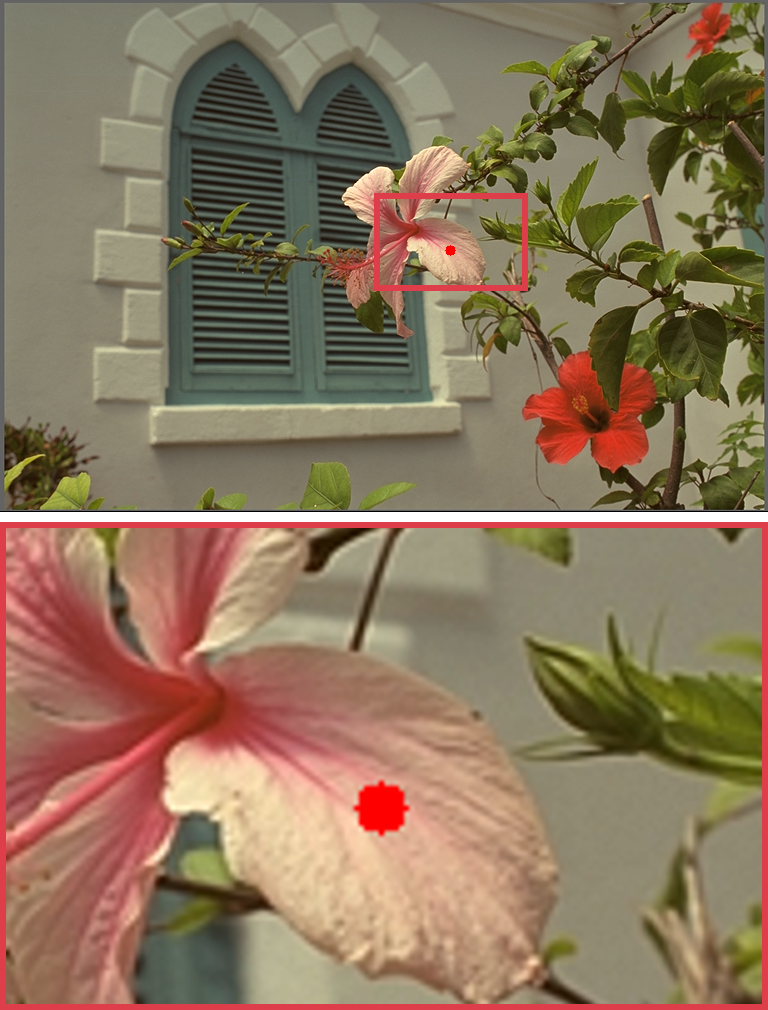}
	\end{subfigure}
	\begin{subfigure}{.24\textwidth}
		\centering
		\includegraphics[width=\textwidth]{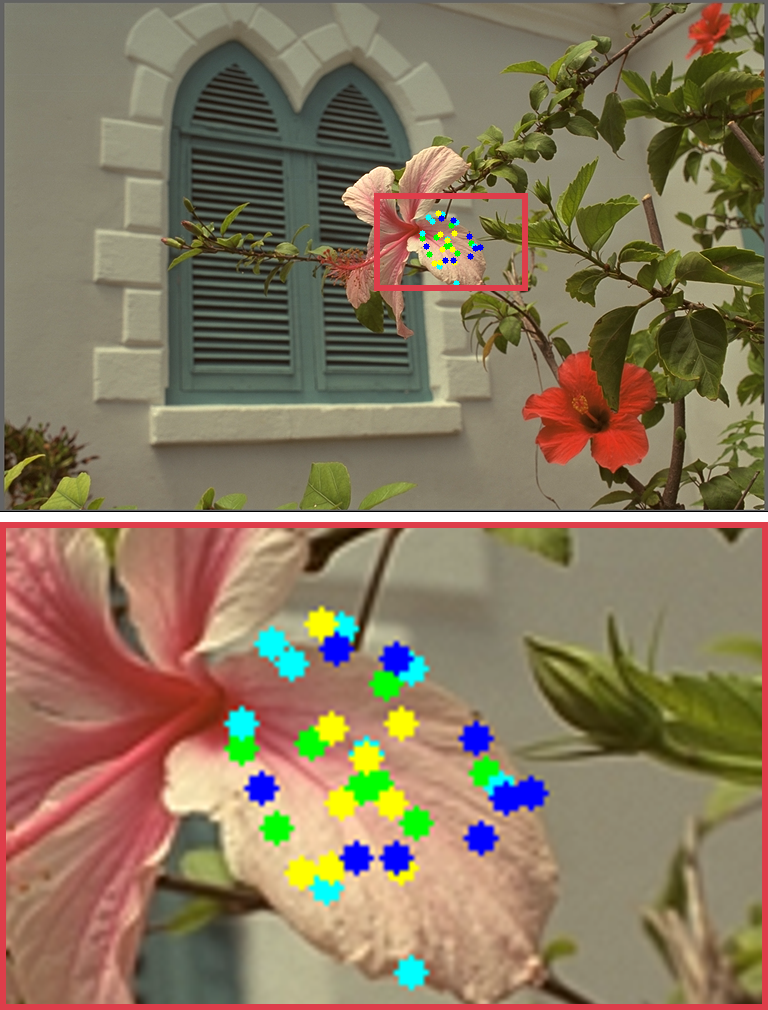}
	\end{subfigure}
	\caption{Visualization of dynamic sampling locations. The left figures are \textit{kodim7} with the red target aggregation point, and the right figures contain the sampling locations of dynamic kernel. Different color denotes different group the points belong to.}
	\label{offset_vis}
\end{figure}

Then we propose to investigate the efficient transform operation containing the LDCN. Since Vision Transformer \cite{vaswani2017attention, liu2021swin, liu2022swin} introduces many powerful modules and can achieve promising performance in computer vision tasks, we test several components in our compression method, including Layer Normalization (LN) \cite{ba2016layer}, Multilayer Perceptron (MLP) \cite{vaswani2017attention} and GELU \cite{hendrycks2016gaussian}. We observe that the RD performance of the compression framework will be improved when using LN behind the LDCN and MLP network. We conduct extensive experiments to explore the best performing network structure with multiple design options (see Section \ref{Design} for more information). Based on our experimental exploration, we design an efficient and stable transformation module named Dynamic Residual Block Group (DRBG). The proposed DRBG is illustrated in Figure~\ref{fig:overview}. Specifically, before conducting LDCN, we follow \cite{wang2022internimage} to use two depth-wise convolutions to generate the content-adaptive kernel offsets and modulations, respectively. Then we use the softmax function to normalize the modulations. As for MLP, we use a simple network to project the feature, and its function can be formulated as:

\begin{equation}
	MLP(\cdot)=Conv_{1\times1}(GELU(Conv_{1\times1}(\cdot))).
	\label{MLP}
\end{equation} 
Also, shortcut connection \cite{he2016deep} is incorporated for the LDCN and the MLP according to the design principles of recent work \cite{liu2021swin, zamir2022restormer, liu2022swin} and our architecture investigation. The stacked dynamic transform modules with a shortcut connection is named Dynamic Residual Block (DRB) in this paper. Furthermore, previous work \cite{cheng2020learned, he2022elic} demonstrate that Residual Block \cite{he2016deep} can boost performance of compression framework by introducing nonlinearity. Therefore, we add a Residual Bottleneck Block (RBB) \cite{he2022elic} after the DRB to further enhance the transform capability. The combination is named into Dynamic Residual Block Group (DRBG).  Detailed architectures of the framework and the DRBG are illustrated in Figure~\ref{fig:overview}.

To further verify the effectiveness of the dynamic kernel, we conduct experiments on visualizing the kernel offsets and the sampling position. Figure~\ref{offset_vis} shows a target aggregation point and corresponding sampling positions. Since the target point is located on the petal, its corresponding dynamic sampling points are located at different positions on the petal to obtain useful spatial information. The phenomenon can demonstrate the dynamic aggregation capacity of our method.
\begin{figure*}[t]
	\centering
	\includegraphics[width=\textwidth]{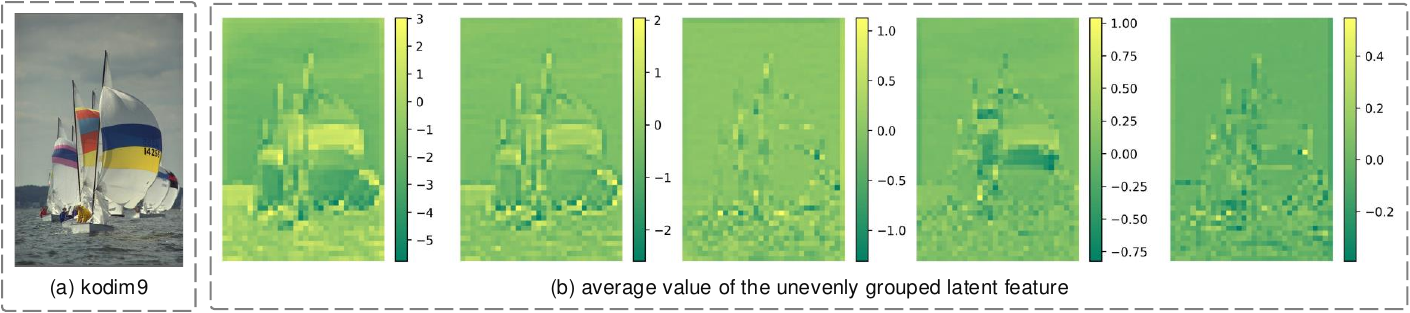}
	\caption{Visualization of the average value of the unevenly grouped latent feature. It can been seen from the figures that the former coding slices have larger symbol magnitudes, and have stronger spatial correlation in the neighborhood.}
	\label{fig:latent_visualization}
\end{figure*}

\begin{figure*}[t]
	\centering
	\includegraphics[width=0.87\textwidth]{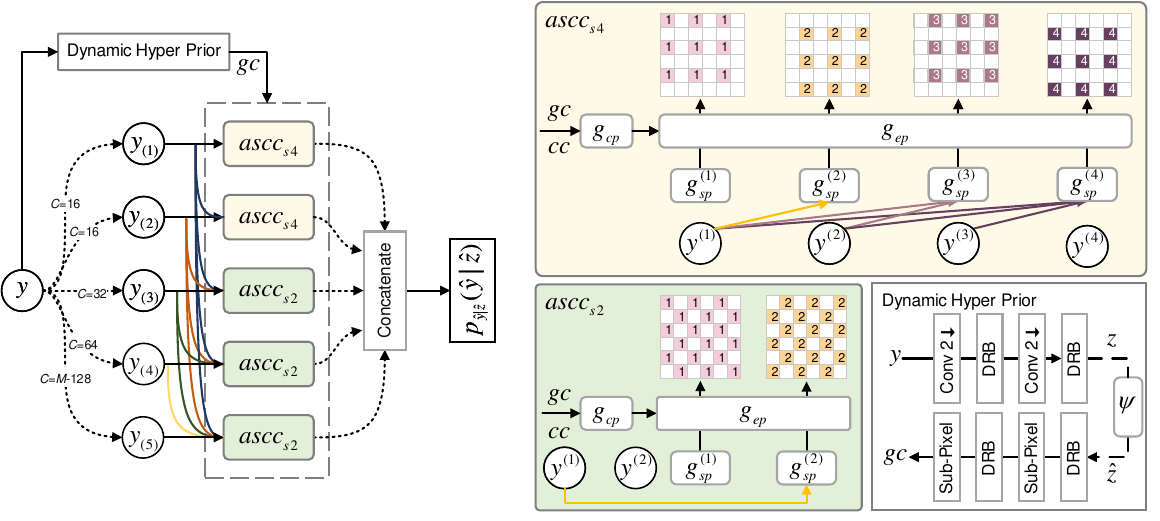}
	\caption{Description of the Asymmetric Spatial-channel Entropy Model. We split the latents $y$ into 5 slices. Every slice has global context $gc$ from hyper-prior. Besides, considering the different spatial correlation in each slice, we use the 4-stage spatial context model to estimate distribution parameters of the first two slices, and we adopt the 2-stage spatial context model to the subsequent slices. The subscript and superscript of $\boldsymbol{y}$ denote the sub part of the latent in channel and spatial dimension respectively.}
	\label{fig:ascem}
\end{figure*}

\subsection{Asymmetric Spatial-channel Entropy Model}\label{ASCEM}
Current progress in efficient entropy models adopts various context information \cite{balle2018variational, minnen2020channel, he2021checkerboard, he2022elic} to facilitate distribution parameter estimation. To utilize all these previous methods' advantages, we first introduce the generalized coarse-to-fine entropy model in Section \ref{framework}. Specifically, we divide the helpful context in image compression into coarse global, channel-wise, and spatial contexts. As Equation \ref{eq:general_EM} described, the coarse global context is obtained by a hyper-prior \cite{balle2018variational} as side information, which has been demonstrated the effectiveness in earlier work \cite{minnen2018joint, cheng2020learned}. As for the fine-grained context, we follow \cite{minnen2020channel, he2021checkerboard} and split the latent representation along the channel and the spatial dimension. Considering the orthogonality of space and channel dimensions, the performance improvement from these two dimensions should also be orthogonal.

For building an accurate yet efficient entropy model, we first introduce dynamic kernel in hyper-prior to generate the enhanced global context with adaptive latent aggregation. The architecture of the Dynamic Hyper Prior is shown in Figure~\ref{fig:ascem}. Then, we follow \cite{he2022elic} and build unevenly grouped channel-wise context. In Figure~\ref{fig:latent_visualization}, we visualize the average latent value of the 5-group model (b), and draw a similar conclusion as in \cite{he2022elic} that the later encoded groups contain less information. From our experimental exploration, squeezing or fusing the former channel groups will cause noticeable performance loss. So we follow \cite{he2022elic} and set the group number to 5, with 16, 16, 32, 64, and 192 channels in the groups.

Furthermore, we notice the decreasing feature spatial correlation during the channel auto-regression process from Figure~\ref{fig:latent_visualization}. This indicates that we can capture the more helpful spatial context in the former groups, and later channel groups are more dependent on the channel context than the spatial context. Following the principle, we introduce the asymmetric channel-spatial entropy model as shown in Figure~\ref{fig:ascem}. From the investigation of our experiments, the first two channel group $y_{(1)}$, $y_{(2)}$ has more explicit spatial neighborhood-dependent properties, so we use a four-stage spatial context model $ascc_{s4}$ (inspired by the multi-stage context in \cite{lu2022high}) to fully leverage the spatial correlation. However, adopting the complex spatial context model in the latter channel groups introduces negligible performance gains and additional computational burden. Therefore, we utilize a two-stage spatial context model $ascc_{s2}$ (inspired by the checkerboard context in \cite{he2021checkerboard}) to reduce the spatial redundancy in the latter channel groups. The entire model, named Asymmetric Spatial-channel Entropy Model, can mine valuable context in a comprehensive manner while maintaining a high inference speed (see Table \ref{tab:BDBR} for more detailed comparisons).

\subsection{Loss Function}\label{sec:loss}
Following most of the previous learned image compression methods \cite{balle2018variational, cheng2020learned}, we use a Lagrangian multiplier to trade off rate and distortion in the loss function. Our framework is trained with the following loss:
\begin{equation}
	\begin{aligned}
		L=&\lambda \cdot D(\boldsymbol{x},\boldsymbol{\hat{x}})+R(\boldsymbol{\hat{y}})+R(\boldsymbol{\hat{z}})\\
		with \quad R(\boldsymbol{\hat{y}})=&\mathbb{E}[-\log_2(p_{\boldsymbol{\hat{y}}|\boldsymbol{\hat{z}}}(\boldsymbol{\hat{y}}|\boldsymbol{\hat{z}}))]\\
		\quad R(\boldsymbol{\hat{z}})=&\mathbb{E}[-\log_2(p_{\boldsymbol{\hat{z}}|\boldsymbol{\psi}}(\boldsymbol{\boldsymbol{\hat{z}}|\boldsymbol{\psi}}))]
	\end{aligned}
	\label{eq:loss}
\end{equation}
in which $R(\cdot)$ denotes the predicted entropy. $\boldsymbol{\hat{y}}$ and $\boldsymbol{\hat{z}}$ are the quantized latent representation and side information. $\boldsymbol{\psi}$ represents the factorized entropy model \cite{balle2016end} to compress the side information. $\lambda$ is the Lagrange multiplier that determines the trade-off between$ R $ and the distortion $D$. To train our model, we use MSE as the distortion function. 

\section{Experimental Results}
\subsection{Datasets and Implementation Details}
\subsubsection{Datasets}
Following the previous work \cite{xie2021enhanced}, we adopt the Flicker2W dataset provided by \cite{liu2020unified}. The dataset contains 20,716 real-world images with various complex textures. To validate the effectiveness of our model and compare performance with other state-of-the-art image compression approaches, we test our models on three benchmarks: the Kodak dataset \cite{kodak}, the CLIC Professional Validation dataset \cite{clic} and the Tecnick dataset \cite{asuni2014testimages}. The resolution of these datasets varies from $768\times512$ to 2K.
\begin{figure*}[t]
	\centering
	\begin{subfigure}{.32\textwidth}
		\centering
		\includegraphics[width=\textwidth]{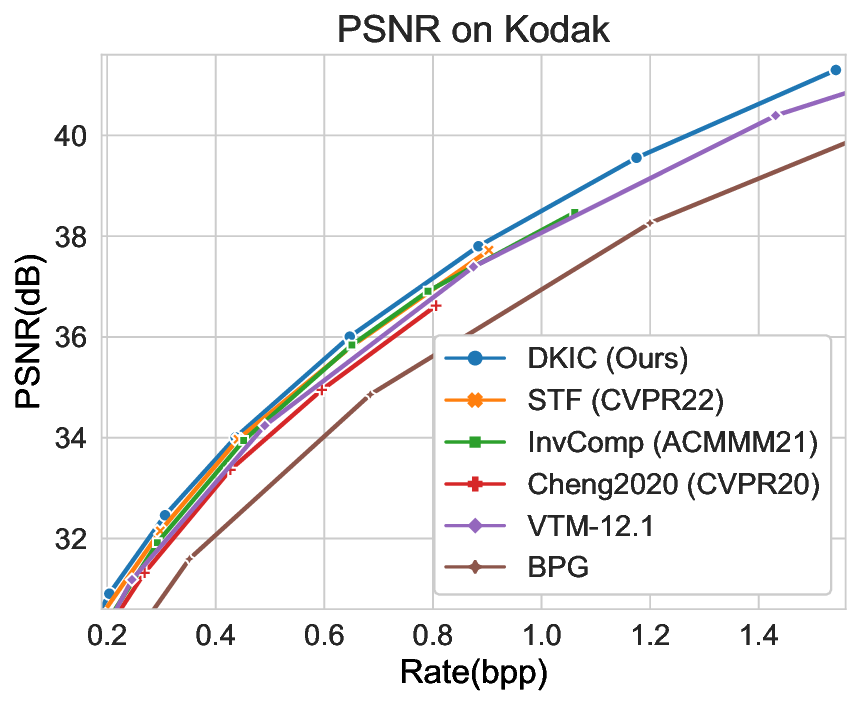}
		\label{Kodak_PSNR}
	\end{subfigure}
	\begin{subfigure}{.32\textwidth}
		\centering
		\includegraphics[width=\textwidth]{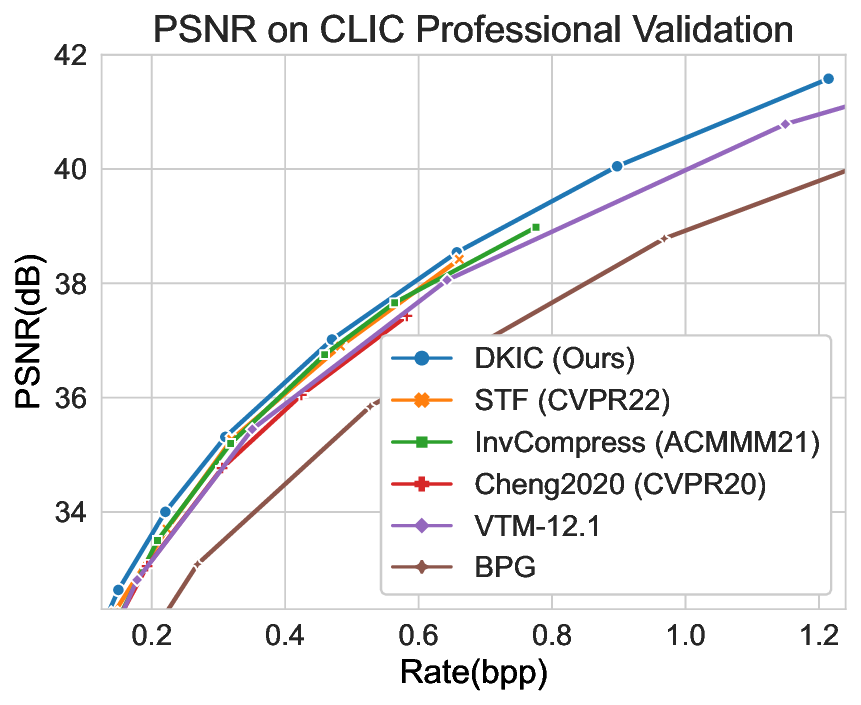}
		\label{CLIC_PSNR}
	\end{subfigure}
	\begin{subfigure}{.32\textwidth}
		\centering
		\includegraphics[width=\textwidth]{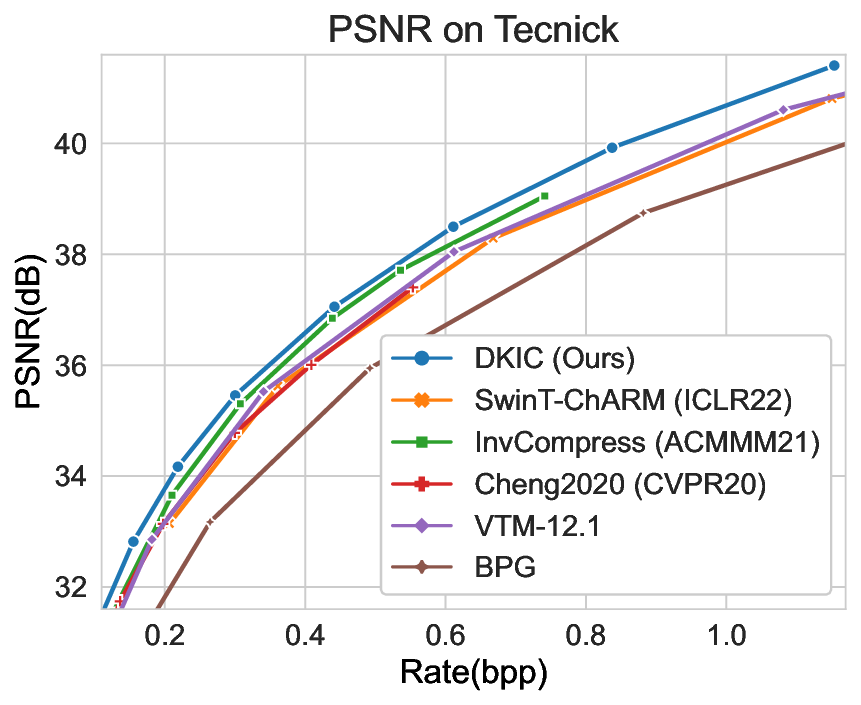}
		\label{Tecnick_PSNR}
	\end{subfigure}
	\caption{The rate-distortion performance of BPG, VTM-12.1, the recent learned image compression approaches and our DKIC on the Kodak, CLIC Professional Validation and Tecnick datasets.}
	\label{RD_curve}
\end{figure*}
\begin{table}[t]
	\centering
	\caption{The comparison results of BDBR (Anchor: VTM-12.1), inference speed and model parameters. The EncT and DecT denote the time cost for encoding and decoding on Kodak, and Para. represents the parameters of each model. A lower BDBR indicates higher RD performance. \textbf{\textcolor{red}{Red}} and \textcolor{blue}{Blue} indicate the best and the second-best performance.}
	\scalebox{0.75}{
		\begin{tabular}{c|ccc|cc|c}
			\toprule[1.2pt]
			Image Codec & Kodak   & CLIC    & Tecnick & \begin{tabular}[c]{@{}c@{}}EncT\\(ms)\end{tabular}  & \begin{tabular}[c]{@{}c@{}}DecT\\ (ms)\end{tabular}& \begin{tabular}[c]{@{}c@{}}Para.\\ (M)\end{tabular} \\ \midrule[1pt]
			BPG444      & 22.39\% & 28.09\% & 23.48\% &-&-&-\\ \midrule
			Cheng2020 \cite{cheng2020learned}   & 5.67\%  & 4.73\%  & 6.38\% & 4395 & 8387&29.6 \\
			InvComp \cite{xie2021enhanced}     & -1.23\% & 	\textcolor{blue}{-3.57\%} & 	\textcolor{blue}{-5.18\%} &4354 & 10641 &47.5\\
			SwinT-ChARM \cite{zhu2022transformerbased} & 	\textcolor{blue}{-3.01\%} & -       & 2.73\% &-&-&60.5\\
			STF \cite{zou2022devil}         & -2.88\% & -3.34\% & -    &159 & 162  &99.8\\
			DKIC (Ours) & \textcolor{red}{ \textbf{-7.68\%}} &\textcolor{red}{ \textbf{ --7.89\% }}& \textcolor{red}{ \textbf{-9.90\%}} &147 &158&53.3\\ \bottomrule[1.2pt]
		\end{tabular}
	}
	\label{tab:BDBR}
\end{table}
\subsubsection{Training settings}
We randomly crop the training dataset with the size of  $256 \times 256$. Using the loss function \ref{eq:loss}, we train six models with different $\lambda $ to control RD performance (Following CompressAI \cite{begaint2020compressai}, we set $\lambda $ = 0.0035, 0.0067, 0.0130, 0.0250, 0.0483, 0.0932, 0.1800 for the model). The batch size is set to 8, and the AdamW optimizer is adopted whose parameters $\beta_{1}$ and $\beta_{2}$ are set as $0.9$ and $0.999$. The learning rate is initialized to $1 \times 10^{-4}$ and decreased to $1 \times 10^{-5}$ after 380 epoch training. The entire network converges after 400 epochs. All experiments are conducted using the PyTorch with NVIDIA RTX 3090 GPUs.

\subsubsection{Test settings}
We test our method on three benchmarks, including Kodak \cite{kodak} with the image size of $768\times 512$, Tecnick test set \cite{asuni2014testimages} with the image size of $1200\times 1200$ and CLIC professional validation dataset \cite{clic} with 2k resolution. We use PSNR to measure the distortion, while bits per pixel (bpp) are used to evaluate bitrates.
\begin{table}[t]
	\centering
	\caption{The BDBR results of ablation studies. A lower BDBR indicates higher RD performance. We vary the anchor in different ablation experiments, and each anchor uses the module in our final method.}
	\scalebox{0.82}{
		
		\begin{tabular}{cccc}
			\toprule[1.2pt]
			\multicolumn{4}{c}{Dynamic Kernel}                                                                                                                                                                                                                                  \\ \midrule[1pt]
			Dynamic Kernel& \multicolumn{1}{|c||}{BDBR}                     & \multicolumn{1}{c|}{Dynamic Kernel Size}                                                  & \multicolumn{1}{c}{BDBR}    \\ \midrule
			\multicolumn{1}{c|}{w/ DRB}                                                                 & \multicolumn{1}{c||}{0.00\%}                   & \multicolumn{1}{c|}{1x1}                                                                  & \multicolumn{1}{c}{5.56\%}  \\ \midrule
			\multicolumn{1}{c|}{Replace DRB with RBB}                                               & \multicolumn{1}{c||}{6.36\%} & \multicolumn{1}{c|}{3x3}                                                                  & \multicolumn{1}{c}{0.00\%}  \\ \midrule
			\multicolumn{1}{c|}{Replace LDCN with Conv}                                                                     & \multicolumn{1}{c||}{8.14\%}                         & \multicolumn{1}{c|}{5x5}                                                                  & \multicolumn{1}{c}{-0.54\%} \\ \midrule[1pt]
			\multicolumn{4}{c}{Asymmetric Entropy Model}                                                                                                                                                                                                                         \\ \midrule[1pt]
			\multicolumn{1}{c|}{\begin{tabular}[c]{@{}c@{}}Spatial Model \\ Arrangement\end{tabular}} & \multicolumn{1}{c||}{BDBR}                     & \multicolumn{1}{c|}{\begin{tabular}[c]{@{}c@{}}Spatial Model \\ Arrangement\end{tabular}} & BDBR                         \\ \midrule
			\multicolumn{1}{c|}{{[}1, 1, 1, 1, 1{]}}                                                  & \multicolumn{1}{c||}{5.43\%}                   & \multicolumn{1}{c|}{{[}4, 4, 2, 2, 2{]}}                                                  & 0.00\%                       \\ \midrule
			\multicolumn{1}{c|}{{[}2, 2, 2, 2, 2{]} \cite{he2022elic}}                                                  & \multicolumn{1}{c||}{2.02\%}                   & \multicolumn{1}{c|}{{[}4, 4, 4, 2, 2{]}}                                                  & -0.04\%                      \\ \midrule
			\multicolumn{1}{c|}{{[}4, 2, 2, 2, 2{]}}                                                  & \multicolumn{1}{c||}{0.65\%}                   & \multicolumn{1}{c|}{{[}4, 4, 4, 4, 4{]}}                                                  & -0.06\%                      \\ 
			\bottomrule[1.2pt]
		\end{tabular}
	}
	\label{tab:ablation}
\end{table}

\subsection{Traditional Codec Evaluation}
In order to compare rate-distortion performance with traditional image compression methods, including BPG \cite{bellard2014bpg} and VTM-12.1 \cite{vvc}, we refer to the setting in \cite{zhu2022transformerbased} to generate the bitstream of BPG and VTM-12.1. Specifically, we obtain BPG from the website\footnote{\href{ https://bellard.org/bpg/}{https://bellard.org/bpg/}}, and we set quality index as 22, 27, 32, 37, 42, 47 for BPG. As for VTM-12.1, we download the reference software from the website\footnote{\href{ https://vcgit.hhi.fraunhofer.de/jvet/VVCSoftware\_VTM/-/tags/VTM-12.1}{ https://vcgit.hhi.fraunhofer.de/jvet/VVCSoftware\_VTM/-/tags/VTM-12.1}}. We follow \cite{zhu2022transformerbased} and use the scripts in CompressAI\footnote{\href{https://github.com/InterDigitalInc/CompressAI/tree/efc69ea24}{https://github.com/InterDigitalInc/CompressAI/tree/efc69ea24}} \cite{begaint2020compressai} to gather VTM results using QP 17, 22, 27, 32, 37, 42, 47.

\subsection{Comparison with Other Methods}

\subsubsection{Comparison Methods}
We compare our method with several recently learned image compression approaches\footnote{We only compare the performance of work with open-sourced codes or available rate-distortion data.}: Cheng2020 \cite{cheng2020learned}, InvComp \cite{xie2021enhanced}, SwinT-ChARM \cite{zhu2022transformerbased} and STF \cite{zou2022devil}. To facilitate comparison, we use \textit{cheng2020-attn} reproduced by CompressAI \cite{begaint2020compressai}, which is comparable to the performance of the original paper. Apart from this, all the rate-distortion data of the comparison methods are provided by their authors. As for the traditional methods BPG and VTM-12.1, we follow the previous work \cite{zhu2022transformerbased} and conduct the experiments of traditional codecs in YUV 4:4:4 colorspace.

\subsubsection{Rate-Distortion Performance}
Figure~\ref{RD_curve} shows the experimental results on the Kodak, CLIC Professional Validation, and Tecnick datasets while taking PSNR as quality measurement. Our method DKIC can outperform traditional image compression methods BPG and VTM-12.1 on every benchmarks. Specifically, at the same bitrate, the proposed method has an average of 0.4dB improvement in PSNR compared to VTM-12.1 on three test datasets, and has an average of 1.4dB improvement in PSNR compared to BPG. DKIC can also achieve superior performance among learned image compression algorithms. We can see from Figure~\ref{RD_curve} that DKIC shows a most promising performance than recent LIC methods, especially in the high bitrate region.
\begin{table}[t]
	\centering
	\caption{The BDBR results of ablation studies on different entropy models. A lower BDBR indicates higher RD performance. AR times denotes auto-regressive times in the entropy model.}
	\scalebox{0.80}{
		
		\begin{tabular}{cccc}
			\toprule[1.2pt]
			Grouping Style& Spatial Context Model& BDBR& AR times                                                                                                                                                                                                          \\ \midrule[1pt]
			w/o slicing \cite{he2021checkerboard}& 2-stage for the latents & 4.16\% & 2\\ \midrule
			w/o slicing \cite{lu2022high}& 4-stage for the latents & 2.79\% & 4 \\ \midrule
			Even (10 slices) \cite{minnen2020channel}  & w/o spatial context model & 0.93\% &10 \\ \midrule
			Even (10 slices)  & 2-stage for each group &0.28\% &20 \\ \midrule
			Even (10 slices)  &4-stage for each group & -0.12\% &40 \\ \midrule
			Asymmetric (Ours) & [4,4,2,2,2] & 0.00\% &14 \\ \bottomrule[1.2pt]
			
		\end{tabular}
	}
	\label{tab:ablation_grouping}
\end{table}

For a more detailed comparison, we also provide the Bjøntegaard Delta Bit-Rate (BDBR) results \cite{bdrate} computed from the rate-distortion curves as the quantitative metric. We set VTM-12.1 as the anchor for performance comparison. The detailed results are shown in Table \ref{tab:BDBR}. Specifically, the proposed method can save 4.51\%, 6.18\%, 7.48\% bitrate compared to VTM-12.1 on Kodak, CLIC, and Tecnick datasets, respectively. STF \cite{zou2022devil} adopted window attention-based Swin Transformer block for transform coding and used channel-wise entropy model for probabilistic prediction. From the BDBR results calculated by PSNR, STF can save 2.88\% bitrate compared to VTM-12.1 on Kodak, and our DKIC can save a further 1.63\% of bits relative to STF. In Table \ref{tab:BDBR}, we also provide the coding time of recent learned image compression methods, including Cheng2020 \cite{cheng2020learned}, InvComp \cite{xie2021enhanced} and STF \cite{zou2022devil}. We perform the speed comparison on the same test environment (single RTX 3090) using the source code provided by these methods. Since Cheng2020 \cite{cheng2020learned} and InvComp \cite{xie2021enhanced} adopted spatial autoregressive entropy model, they must encode and decode each element of the latents sequentially. In that case, this method is not practical with the non-parallel operations. It takes 8 to 10 seconds for them to decode a single $768\times512$ size image. DKIC uses the asymmetric spatial-channel entropy model with controlled complexity, and it achieves the best RD performance among the comparison methods while maintaining the highest coding efficiency.

\begin{table}[t]
	\centering
	\caption{The BDBR results of ablation studies on dynamic residual block architecture design. A lower BDBR indicates higher RD performance.}
	\scalebox{1}{
		\begin{tabular}{cccc}
			\toprule[1.2pt]
			\multicolumn{3}{c|}{Dynamic Residual Block Architecture Design}                                                                                                                                                                                        & BDBR                         \\ \midrule
			\multicolumn{3}{c|}{\begin{subfigure}{.3\textwidth}
					\centering
					\includegraphics[width=\textwidth]{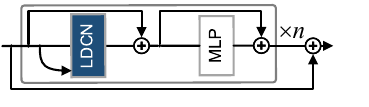}
			\end{subfigure}}                                                                                                                                                                                                                 &         \raisebox{3\height}{1.22\%}\\ \midrule
			\multicolumn{3}{c|}{\begin{subfigure}{.3\textwidth}
					\centering
					\includegraphics[width=\textwidth]{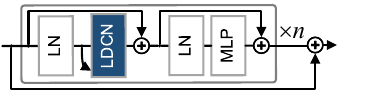}
			\end{subfigure}}  &\raisebox{3\height}{0.87\%}                            \\ \midrule
			\multicolumn{3}{c|}{\begin{subfigure}{.3\textwidth}
					\centering
					\includegraphics[width=\textwidth]{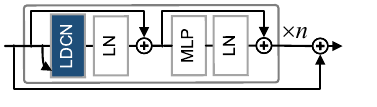}
			\end{subfigure}}   &   \raisebox{3\height}{0.00\%} \\ \midrule
			\multicolumn{3}{c|}{\begin{subfigure}{.3\textwidth}
					\centering
					\includegraphics[width=\textwidth]{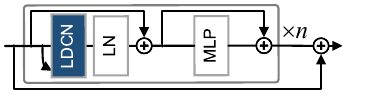}
			\end{subfigure}}   &  \raisebox{3\height}{0.33\%} \\ \midrule
			\multicolumn{3}{c|}{\begin{subfigure}{.3\textwidth}
					\centering
					\includegraphics[width=\textwidth]{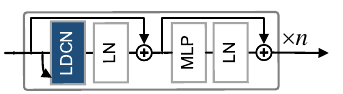}
			\end{subfigure}}   &  \raisebox{3\height}{1.67\%} \\\bottomrule[1.2pt]
		\end{tabular}
	}
	\label{tab:DRB_ablation}
\end{table}
\begin{figure*}[t!]
	\centering
	\includegraphics[width=0.91\textwidth]{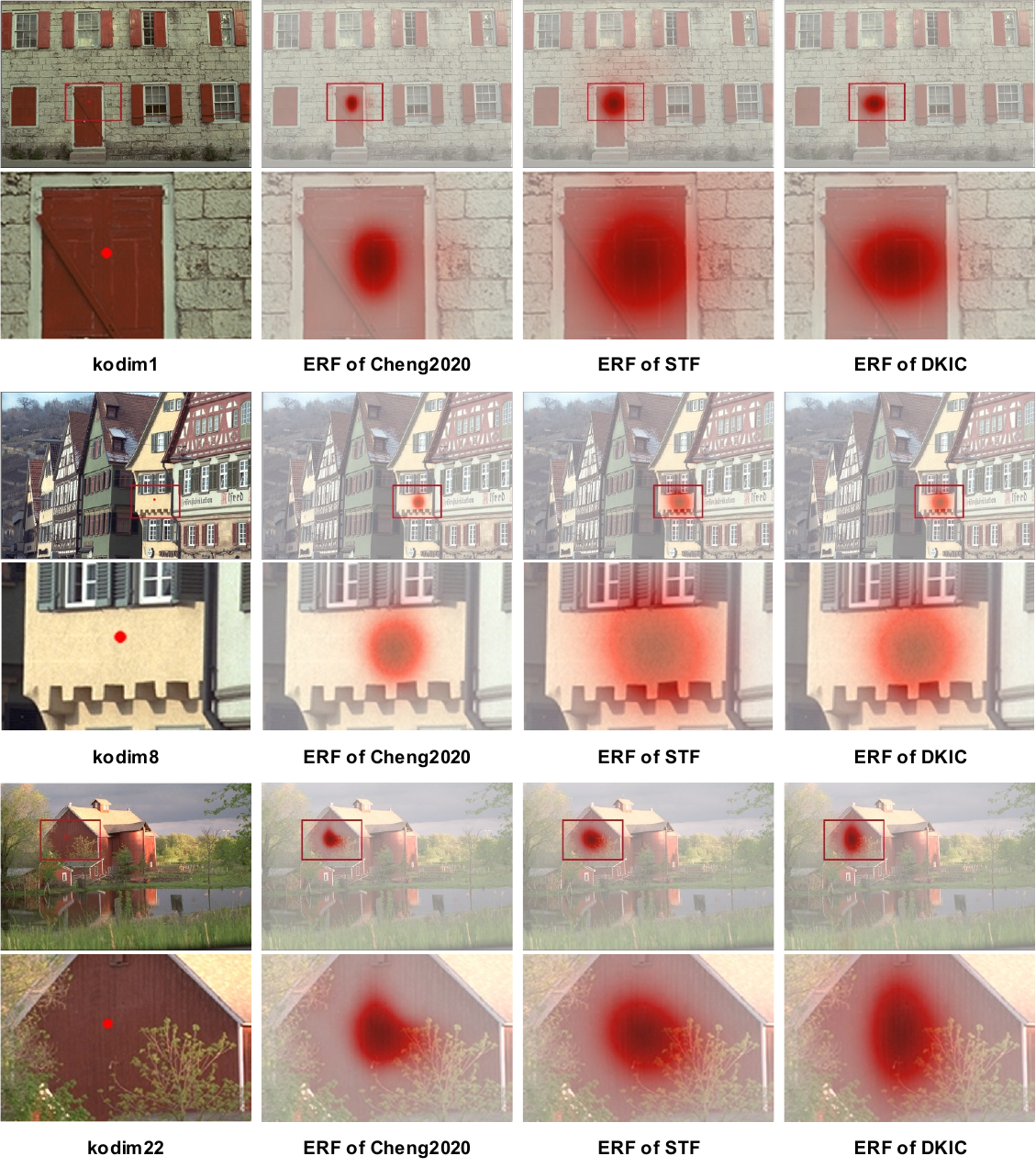}
	\caption{Visualization of the effective receptive fields, we choose Cheng2020\cite{cheng2020learned} and STF\cite{zou2022devil} as the representative methods using ordinary kernel and window-based attention, respectively. \textcolor{red}{The red point} denotes the target point.}
	\label{fig:erf}
\end{figure*}
\subsection{Ablation Studies and Performance Analysis}
\begin{figure*}[t]
	\centering
	\includegraphics[width=0.94\textwidth]{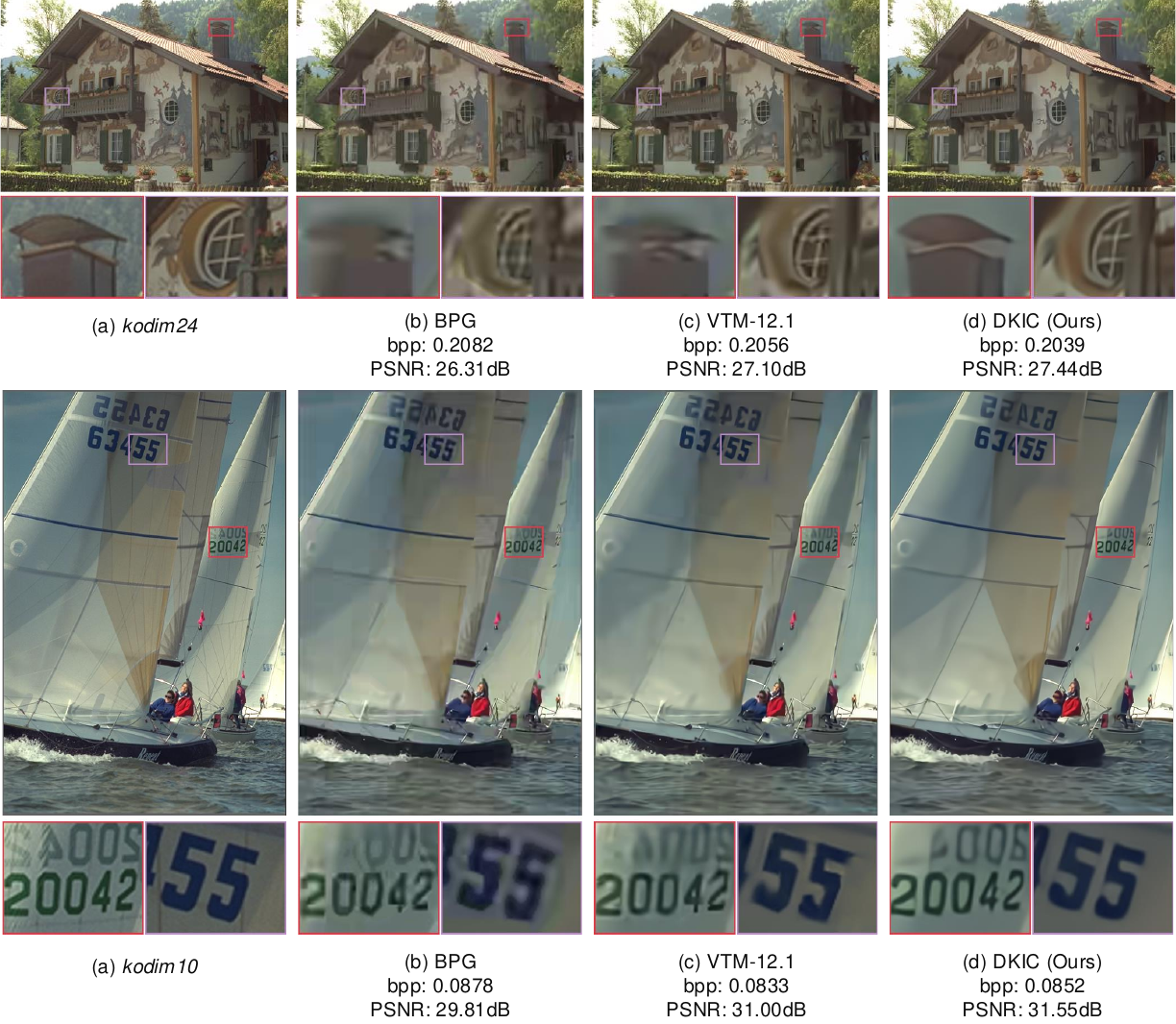}
	\caption{The visual results of traditional image compression methods BPG, VTM-12.1 and our DKIC.}
	\label{fig:subjective_comparison}
\end{figure*}
\begin{figure*}[t!]
	\centering
	\includegraphics[width=0.94\textwidth]{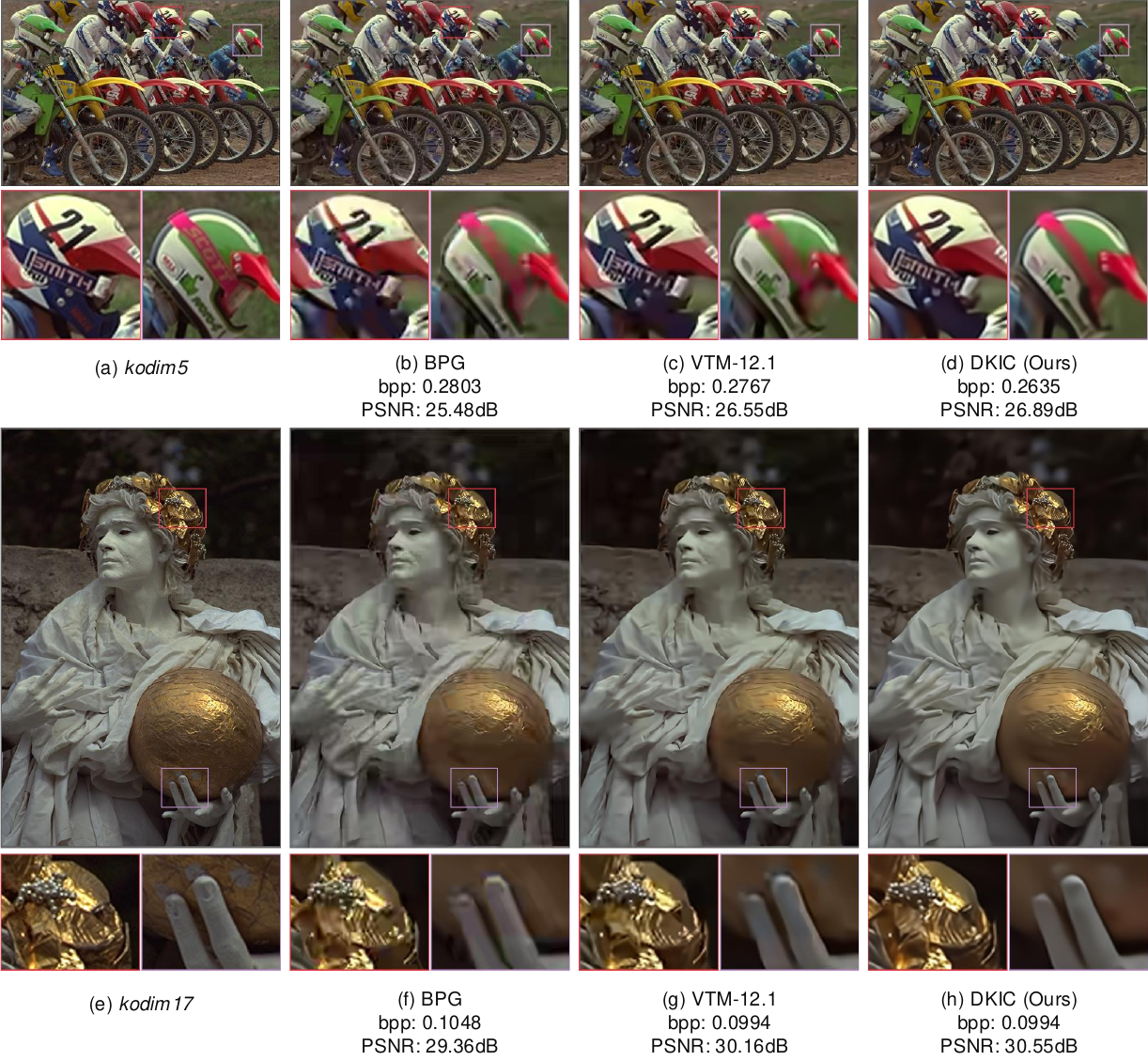}
	\caption{The visual results of traditional image compression methods BPG, VTM-12.1 and our DKIC.}
	\label{fig:subjective_comparison2}
\end{figure*}
\subsubsection{Dynamic Kernel}
To verify the dynamic kernel's effectiveness, we first conduct experiments by replacing the dynamic residual block with the residual bottleneck block. Without the adaptive spatial aggregation capacity, the rate will increase by about 6.36\% at the same PSNR. We also directly replace the LDCN with ordinary convolution, and the rate will increase by about 8.14\% at the same PSNR. Then we vary the dynamic kernel size to investigate the aggregation capacity of different kernel sizes. From Table \ref{tab:ablation}, the bits will increase by 5.56\% when we replace the 3$\times$3 dynamic kernel with 1$\times$1 kernel. Then we replace the 3$\times$3 dynamic kernel with 5$\times$5 kernel. The bits will save about 0.54\%. We can infer from the experiments that as the size of the dynamic kernel becomes larger, the dynamic aggregation capacity increases. However, the improvement in model performance decreases significantly when the dynamic aggregation capacity reaches a bottleneck. Considering the high computation complexity and GPU memory usage caused by 5$\times$5 dynamic kernel, we choose 3$\times$3 size kernel in our model.

\subsubsection{Asymmetric Spatial-channel Entropy Model}
To evaluate the effectiveness of our proposed Asymmetric Spatial-channel Entropy Model, we conduct extensive experiments on changing the stage number of the spatial context model in each split slice. Since we divide the latent representation into five slices, we denote the list in Table \ref{tab:ablation} as the spatial context stage for each slice. For example, [2, 2, 2, 2, 2] represents that each slice uses a two-stage spatial context. As shown in Figure~\ref{fig:latent_visualization}, the former encoded slices have a stronger spatial correlation in the neighborhood, so we prioritize using a four-stage spatial context model for slices encoded first. Moreover, the complex spatial context used in the later slices brings negligible gain, so our method chooses the Asymmetric Spatial-channel Entropy Model with [4, 4, 2, 2, 2] spatial context arrangement.

\subsubsection{Entropy Model Switching}
We have added additional experimental results to further varify the superior of the proposed entropy model. For example, the results of model comparisons using two-stage\cite{he2021checkerboard} versus four-stage spatial contexts\cite{lu2022high} without latent grouping. We have tabulated the results in Table~\ref{tab:ablation_grouping}, and the proposed entropy model can achieve promising RD performance with acceptable auto-regressive time.

\subsubsection{Design of Dynamic Residual Block}\label{Design}
We also conduct experiments to explore the highest-performance dynamic residual block design options. We set up four comparison methods by adjusting the position of Layer Normalization (LN) and whether to remove LN. The difference of architectures are provided in Table \ref{tab:DRB_ablation}. From the results, we can see that using LN behind the LDCN and the MLP achieves the best performance. Therefore, we choose the post-norm architecture in our implementation. This choice also coincides with the conclusion in many methods \cite{liu2021swin, wang2022internimage}.

\subsubsection{Model Complexity}
\label{com}
DKIC has about 53.3M parameters while the comparison methods InvComp \cite{xie2021enhanced}, SwinT-ChARM \cite{zhu2022transformerbased} and STF \cite{zou2022devil} have 47.5M, 60.5M, and 99.8M parameters respectively. We record the coding time and RD performance of different methods on the Kodak dataset, and the detailed experimental results are provided in Table \ref{tab:BDBR}. It takes DKIC around 150ms to encode or decode a $768\times512$ size image using a single RTX 3090. STF \cite{zou2022devil} has a similar coding speed, but our method can achieve better compression performance. Cheng2020 \cite{cheng2020learned} and InvComp \cite{xie2021enhanced} used joint autoregressive entropy model \cite{minnen2018joint}, so it takes much longer time to compress. DKIC can achieve the most promising RD performance while maintaining satisfying model complexity.

\subsection{Effective Receptive Fields}

To further demonstrate the effectiveness of our proposed Dynamic Kernel, we present more effective receptive fields (ERF) visualization of different methods in Figure~\ref{fig:erf}. We continue to choose Cheng2020 \cite{cheng2020learned}, STF\footnote{\href{ https://github.com/Googolxx/STF}{ https://github.com/Googolxx/STF}} \cite{zou2022devil} and our DKIC as representative methods of using the ordinary kernel, window-based attention and dynamic kernel, respectively. As for generating effective receptive fields, we use local attribution maps (LAM) \cite{gu2021interpreting} to find the input pixels that strongly influence the reconstruction results. The LAM employs path integral gradients to conduct attribution analysis. We choose three images \textit{kodim1}, \textit{kodim8} and \textit{kodim22} from the Kodak dataset. As an ordinary kernel-based method, Cheng2020 has a relatively small ERF since its kernel size is fixed. Besides, as shown in Figure~\ref{fig:erf} about \textit{kodim22}, it may aggregate information from unrelated regions. STF uses window-based attention to achieve transform coding. We can see that STF has a much larger ERF than Cheng2020 with the merit of the large window and the shifted-window mechanism. However, it tends to aggregate spatial information from unrelated regions too. For example, when we set the central area of the door as the target point, it will collect many irrelevant feature points outside the door for information aggregation. From these three examples, we can conclude that DKIC has fairly large receptive fields under the premise of sampling highly relevant information. From the ERF of DKIC on \textit{kodim22}, DKIC tends to aggregate the pixels associated with the red wall and tries to avoid sampling unrelated pixels such as branches, so its receptive field shape is oval-like. 

\subsection{Visualization Results}
Figure~\ref{fig:subjective_comparison} and Figure~\ref{fig:subjective_comparison2} shows the visual quality comparison of the traditional compression methods BPG \cite{bellard2014bpg}, VTM-12.1 \cite{vvc}, and our DKIC. We choose $kodim24$, $kodim10$, $kodim5$ and $kodim17$ with complex textures from the Kodak dataset. From the reconstruction images, it can be seen that our method can achieve higher subject quality with clearer texture and more accurate object contours while maintaining better rate-distortion performance.

\section{Conclusions}
In this paper, we present a learned image compression method with dynamic kernel-based adaptive spatial aggregation. The dynamic kernel can generate content-adaptive kernel offset, so the transform capacity is also significantly improved by the adaptive spatial aggregation. We design a new image compression framework by combining the dynamic kernel-based transform coding and nonlinearity from residual bottleneck block. Besides, we define a generalized coarse-to-fine entropy model considering the global, channel-wise, and spatial context. Then according to the latent distribution of channel-wise context, we propose the asymmetric spatial-channel entropy model to reduce statistical redundancy while maintaining high coding efficiency. 

Experimental results demonstrate that our method can obtain better RD performance than the traditional image compression method VTM-12.1 and other state-of-the-art image compression approaches.

\bibliographystyle{IEEEtran}
\bibliography{ms}

\end{document}